# Analytical Solution for BPM Reading the Multi-bunch Average in Booster

Xi Yang

*Fermi National Accelerator Laboratory*

Box 500, Batavia IL 60510

**Abstract**

The BPM system in Booster only can provide the beam position from the average of about 15 bunches due to the electronic limitation. Numerically calculating the difference made by this average can nearly give all the information, which is needed for extrapolating the BPM data. It is still useful to derive the analytical expression for the BPM reading, which can be helpful in getting the insight for the BPM system optimization.

## Introduction

The simulation for comparing the difference between BPM reading one bunch and the average of multi-bunches has already been done,[1] except for the detail derivation of the analytical solution for BPM reading the multi-bunch average. It is worth the effort to go through all the algebra in order to get an analytical solution. Afterwards, one can gain some insights from the analytical solution, and also have a better understanding of the BPM data for the purpose of diagnostics and experimental parameter optimization.

**Derivation**

A BPM reading at the $m^{th}$ turn ($x$) can be written as the summation of three different terms, as shown in eq.1.[2]

$$x(m) = x_{co} + x_D(m) + x_\beta(m). \tag{1}$$

$x_{co}$ represents the close orbit, $x_D$ represents the displacement due to the none-zero dispersion ($D$), and $x_\beta$ is the displacement due to the betatron motion.

In the situation that the BPM reads one bunch, $x_D$ is

$$x_D(m) = D_x \times \delta \times \cos(2\pi Q_s m + \phi_0). \tag{2}$$

$Q_s$ is the synchrotron tune, and $\phi_0$ is the phase of the synchrotron motion at the BPM right after the excitation, which can be set to zero since the phase advance of the synchrotron motion is small within one Booster turn. $\delta$ is

$$\delta = \Delta p / p. \tag{3}$$

$x_\beta$ is

$$x_\beta(m) = \hat{x} \times \cos(2\pi Q_x m + \frac{\xi_x \times \delta \times \sin(2\pi Q_s m + \phi_0)}{Q_s} + \phi'_0). \tag{4}$$

$\hat{x}$ is the amplitude of the betatron motion at the BPM, $Q_x$ is the betatron tune, and $\xi_x$ is the chromaticity. $\phi'_0$ is the phase of the betatron motion at the BPM on the $0^{th}$ turn; it depends upon the phase advance between the place where the betatron motion is excited and the BPM position. Here, we set it to zero.

In the situation that the BPM reads the average of N bunches, $x_{co}$ is the same as before, $x_D$ and $x_\beta$ become the average of N adjacent bunches. Here, N is an integer, and it is about 15 for Booster. The betatron phase difference between the $n^{th}$ bunch and the $1^{st}$ bunch is $\frac{2\pi Q_x}{h} \cdot (n-1)$, and their synchrotron phase difference is $\frac{2\pi Q_s}{h} \cdot (n-1)$. Here, $h$ is the harmonic number. $x_D(m,N)$ is derived as following:

$$x_D(m,N) = \frac{1}{N} \times \sum_{n=1}^{N} D \cdot \delta \cdot \cos\left[2\pi Q_s m + \frac{2\pi Q_s}{h}(n-1)\right]$$

$$= \frac{D\delta}{N} \cdot \sum_{n=1}^{N} \frac{\left\{\exp\left[j\left(2\pi Q_s m + \frac{2\pi Q_s}{h}(n-1)\right)\right] + \exp\left[-j\left(2\pi Q_s m + \frac{2\pi Q_s}{h}(n-1)\right)\right]\right\}}{2}$$

$$= \frac{D\delta}{2N} \cdot \exp[j(2\pi Q_s m)] \cdot \sum_{n=1}^{N} \exp\left[j\left(\frac{2\pi Q_s}{h}(n-1)\right)\right]$$

$$+ \frac{D\delta}{2N} \cdot \exp[-j(2\pi Q_s m)] \cdot \sum_{n=1}^{N} \exp\left[-j\left(\frac{2\pi Q_s}{h}(n-1)\right)\right]$$

$$= \frac{D\delta}{2N}\left\{\exp[j(2\pi Q_s m)] \cdot \frac{\exp\left[j\left(\frac{2\pi Q_s}{h}N\right)\right]-1}{\exp\left[j\left(\frac{2\pi Q_s}{h}\right)\right]-1} + \exp[-j(2\pi Q_s m)] \cdot \frac{\exp\left[-j\left(\frac{2\pi Q_s}{h}N\right)\right]-1}{\exp\left[-j\left(\frac{2\pi Q_s}{h}\right)\right]-1}\right\}$$

$$= \frac{D\delta}{2N}\exp[j(2\pi Q_s m)] \cdot \exp\left[j\left(\frac{2\pi Q_s}{h}\left(\frac{N-1}{2}\right)\right)\right] \cdot \frac{\sin\left(\frac{2\pi Q_s \cdot (N/2)}{h}\right)}{\sin\left(\frac{2\pi Q_s \cdot (1/2)}{h}\right)}$$

$$+ \frac{D\delta}{2N}\exp[-j(2\pi Q_s m)] \cdot \exp\left[-j\left(\frac{2\pi Q_s}{h}\left(\frac{N-1}{2}\right)\right)\right] \cdot \frac{\sin\left(\frac{2\pi Q_s \cdot (N/2)}{h}\right)}{\sin\left(\frac{2\pi Q_s \cdot (1/2)}{h}\right)}$$

$$= \frac{D\delta}{N} \cdot \sin c\left(\frac{Q_s}{h} \cdot N\right) \cdot \cos\left(2\pi Q_s m + \frac{2\pi Q_s}{h}\left(\frac{N-1}{2}\right)\right).$$

(5)

Here, $\sin c(xN) = \frac{\sin(\pi xN)}{\sin(\pi x)}$.

$x_\beta(m,N)$ can be calculated using eq.6(a):

$$x_\beta(m,N) = \frac{\hat{x}}{N} \cdot \sum_{n=1}^{N} \cos\left(2\pi Q_x m + \frac{\xi_x \delta}{Q_s} \cdot \sin\left(2\pi Q_s m + \frac{2\pi Q_s}{h}(n-1)\right) + \frac{2\pi Q_x}{h}(n-1)\right). \quad (6(a))$$

Since the number of bunches to be averaged by the BPM in Booster is about 15 and the synchrotron phase covered by the bunch train ($N=15$) is not more than 4° (at injection),

$\frac{2\pi Q_s}{h}(n-1)$ can be neglected in eq.6(a). Therefore, eq.6(a) can be simplified as following:

$$x_\beta(m,N) = \frac{\hat{x}}{N} \cdot \sum_{n=1}^{N} \cos\left(2\pi Q_x m + \frac{\xi_x \delta}{Q_s} \cdot \sin(2\pi Q_s m) + \frac{2\pi Q_x}{h}(n-1)\right)$$

$$= \frac{\hat{x}}{2N} \cdot \sum_{n=1}^{N} \exp\left(j\left(2\pi Q_x m + \frac{\xi_x \delta}{Q_s} \cdot \sin(2\pi Q_s m) + \frac{2\pi Q_x}{h}(n-1)\right)\right)$$

$$+ \frac{\hat{x}}{2N} \cdot \sum_{n=1}^{N} \exp\left(-j\left(2\pi Q_x m + \frac{\xi_x \delta}{Q_s} \cdot \sin(2\pi Q_s m) + \frac{2\pi Q_x}{h}(n-1)\right)\right)$$

$$= \frac{\hat{x}}{2N} \exp\left(j\left(2\pi Q_x m + \frac{\xi_x \delta}{Q_s} \cdot \sin(2\pi Q_s m)\right)\right) \cdot \sum_{n=1}^{N} \exp\left(j\left(\frac{2\pi Q_x}{h}(n-1)\right)\right)$$

$$+ \frac{\hat{x}}{2N} \exp\left(-j\left(2\pi Q_x m + \frac{\xi_x \delta}{Q_s} \cdot \sin(2\pi Q_s m)\right)\right) \cdot \sum_{n=1}^{N} \exp\left(-j\left(\frac{2\pi Q_x}{h}(n-1)\right)\right)$$

$$= \frac{\hat{x}}{2N} \exp\left(j\left(2\pi Q_x m + \frac{\xi_x \delta}{Q_s} \cdot \sin(2\pi Q_s m)\right)\right) \cdot \exp\left(j\left(\frac{2\pi Q_x}{h}\left(\frac{N-1}{2}\right)\right)\right) \cdot \text{sinc}\left(\frac{Q_x}{h}N\right)$$

$$+ \frac{\hat{x}}{2N} \exp\left(-j\left(2\pi Q_x m + \frac{\xi_x \delta}{Q_s} \cdot \sin(2\pi Q_s m)\right)\right) \cdot \exp\left(-j\left(\frac{2\pi Q_x}{h}\left(\frac{N-1}{2}\right)\right)\right) \cdot \text{sinc}\left(\frac{Q_x}{h}N\right)$$

$$= \frac{\hat{x}}{N} \cdot \text{sinc}\left(\frac{Q_x}{h}N\right) \cdot \cos\left(2\pi Q_x m + \frac{\xi_x \delta}{Q_s} \cdot \sin(2\pi Q_s m) + \frac{2\pi Q_x}{h}\left(\frac{N-1}{2}\right)\right). \quad (6(b))$$

Finally, the BPM reading from the average of $N$ bunches can be written as a function of time ($t$) by replacing $m$ in eqs.5 and 6(b) with $t$ and timing a comb function $\sum_{m=-\infty}^{\infty} \delta(t - mT - \hat{\tau}\cos(2\pi Q_s t - \varphi))$, [3,4] as shown in eq.7.

$$x(t,N) = \begin{pmatrix} x_{co} + \frac{D\delta}{N} \cdot \text{sinc}\left(\frac{Q_s}{h} \cdot N\right) \cdot \cos\left(2\pi Q_s t + \frac{2\pi Q_s}{h}\left(\frac{N-1}{2}\right)\right) \\ + \frac{\hat{x}}{N} \cdot \text{sinc}\left(\frac{Q_x}{h}N\right) \cdot \cos\left(2\pi Q_x t + \frac{\xi_x \delta}{Q_s} \cdot \sin(2\pi Q_s t) + \frac{2\pi Q_x}{h}\left(\frac{N-1}{2}\right)\right) \end{pmatrix} \quad (7)$$

$$\cdot \sum_{m=-\infty}^{\infty} \delta(t - mT - \hat{\tau}\cos(2\pi Q_s t - \varphi)).$$

Here, $\hat{\tau}$, $2\pi Q_s$, and $\varphi$ are the amplitude, the angular frequency, and the phase of the synchrotron motion respectively, and $T$ is the revolution period.

Since we know that whenever two functions $f_1(t)$ and $f_2(t)$ are multiplied in time domain, they get convolved in the frequency domain according to the convolution theorem, eq.7 can be rewritten as $x(t,N) = f_1(t) \cdot f_2(t)$. Here,

$$f_1(t) = x_{co} + \frac{D\delta}{N} \cdot \text{sinc}\left(\frac{Q_s}{h} \cdot N\right) \cdot \cos\left(2\pi Q_s t + \frac{2\pi Q_s}{h}\left(\frac{N-1}{2}\right)\right)$$
$$+ \frac{\hat{x}}{N} \cdot \text{sinc}\left(\frac{Q_x}{h} N\right) \cdot \cos\left(2\pi Q_x t + \frac{\xi_x \delta}{Q_s} \cdot \sin(2\pi Q_s t) + \frac{2\pi Q_x}{h}\left(\frac{N-1}{2}\right)\right), \tag{8}$$

$$f_2(t) = \sum_{m=-\infty}^{\infty} \delta(t - mT - \hat{\tau}\cos(2\pi Q_s t - \varphi)). \tag{9}$$

The convolution is defined as:

$$F_1(\omega) * F_2(\omega) = \int_{-\infty}^{\infty} F_1(\omega') \cdot F_2(\omega - \omega') \cdot d\omega'. \tag{10}$$

For the Fourier transform (FT) of $f_2(t)$ with the aid of eq.11,[4]

$$\exp(-jx\cos\varphi) = \sum_{l=-\infty}^{\infty} j^{-l} J_l(x) \exp(jl\varphi), \tag{11}$$

we can rewrite $f_2(t)$ and get the FT ($F_2(\omega)$) according to eq.12.

$$f_2(t) = \sum_{m=-\infty}^{\infty} \delta(t - mT - \hat{\tau}\cos(2\pi Q_s t - \varphi)) = \frac{\omega_0}{2\pi} \sum_{k=-\infty}^{\infty} \exp(jk(\omega_0 t - \omega_0 \hat{\tau}\cos(\omega_s t - \varphi)))$$
$$= \frac{\omega_0}{2\pi} \sum_{k=-\infty}^{\infty} \sum_{l=-\infty}^{\infty} j^{-l} \cdot J_l(k\omega_0 \hat{\tau}) \cdot \exp(j(k\omega_0 + l\omega_s)t) \cdot \exp(jl\varphi) \xrightarrow{\text{Fourier transform}} \tag{12}$$
$$F_2(\omega) = \frac{\omega_0}{2\pi\sqrt{2\pi}} \sum_{k=-\infty}^{\infty} \sum_{l=-\infty}^{\infty} j^{-l} \cdot J_l(k\omega_0 \hat{\tau}) \cdot \exp(jl\varphi) \cdot \delta(\omega - (k\omega_0 + l\omega_s)).$$

Here, $\omega_0 = 2\pi/T$, $\omega_s = 2\pi Q_s$. For the FT of $f_1(t)$, we rewrite $f_1(t)$ as $f1(x) + f2(x)$.

$$f1(x) = x_{co} + \frac{D\delta}{N} \cdot \text{sinc}\left(\frac{Q_s}{h} \cdot N\right) \cdot \cos\left(\omega_s t + \frac{\omega_s}{h}\left(\frac{N-1}{2}\right)\right),$$

$$f2(x) = \frac{\hat{x}}{N} \cdot \text{sinc}\left(\frac{Q_x}{h} N\right) \cdot \cos\left(\omega_x t + \frac{\xi_x \delta}{Q_s} \cdot \sin(\omega_s t) + \frac{\omega_x}{h}\left(\frac{N-1}{2}\right)\right)$$
$$= A(\exp(j(\omega_x t + \hat{\tau}_c \sin(\omega_s t) + \theta)) + \exp(-j(\omega_x t + \hat{\tau}_c \sin(\omega_s t) + \theta)))$$
$$= A\left(\exp\left(j\left(\omega_x t - (-\hat{\tau}_c)\cos\left(\omega_s t - \frac{\pi}{2}\right) + \theta\right)\right) + \exp\left(j\left(-\omega_x t - \hat{\tau}_c \cos\left(\omega_s t - \frac{\pi}{2}\right) - \theta\right)\right)\right)$$

Where $A = \frac{\hat{x}}{2N} \cdot \text{sinc}\left(\frac{Q_x}{h} N\right)$, $\omega_x = 2\pi Q_x$, $\hat{\tau}_c = \frac{\xi_x \delta}{Q_s}$, $\theta = \frac{\omega_x}{h}\left(\frac{N-1}{2}\right)$.

We get the FT of $f_1(t)$, as shown by eq.13.

$$F1(\omega) = \frac{1}{\sqrt{2\pi}} \cdot x_{co} \cdot \delta(\omega) + \frac{1}{2\sqrt{2\pi}} \cdot \frac{D\delta}{N} \cdot \text{sinc}\left(\frac{Q_s}{h} \cdot N\right)$$
$$\cdot \left(\exp\left(j\left(\frac{\omega_s}{h}\left(\frac{N-1}{2}\right)\right)\right) \cdot \delta(\omega - \omega_s) + \cdot \exp\left(-j\left(\frac{\omega_s}{h}\left(\frac{N-1}{2}\right)\right)\right) \cdot \delta(\omega + \omega_s)\right). \quad (13)$$

Also, the FT of $f_2(t)$ is obtained using the similar method with eq.12.[4]

$$F2(\omega) = \frac{A}{\sqrt{2\pi}} \cdot \exp(j\theta) \cdot \sum_{l=-\infty}^{\infty} j^{-l} \cdot J_l(-\hat{\tau}_c) \cdot \delta(\omega - (\omega_x + l\omega_s)) \cdot \exp\left(j \cdot l \cdot \frac{\pi}{2}\right)$$
$$+ \frac{A}{\sqrt{2\pi}} \cdot \exp(-j\theta) \cdot \sum_{l=-\infty}^{\infty} j^{-l} \cdot J_l(\hat{\tau}_c) \cdot \delta(\omega - (-\omega_x + l\omega_s)) \cdot \exp\left(j \cdot l \cdot \frac{\pi}{2}\right). \quad (14)$$

Finally, the FT of $f_1(t)$ is

$$F_1(\omega) = \frac{1}{\sqrt{2\pi}} \cdot x_{co} \cdot \delta(\omega) + \frac{1}{2\sqrt{2\pi}} \cdot \frac{D\delta}{N} \cdot \text{sinc}\left(\frac{Q_s}{h} \cdot N\right)$$
$$\cdot \left(\exp\left(j\left(\frac{\omega_s}{h}\left(\frac{N-1}{2}\right)\right)\right) \cdot \delta(\omega - \omega_s) + \cdot \exp\left(-j\left(\frac{\omega_s}{h}\left(\frac{N-1}{2}\right)\right)\right) \cdot \delta(\omega + \omega_s)\right)$$
$$+ \frac{A}{\sqrt{2\pi}} \cdot \exp(j\theta) \cdot \sum_{l=-\infty}^{\infty} j^{-l} \cdot J_l(-\hat{\tau}_c) \cdot \delta(\omega - (\omega_x + l\omega_s)) \cdot \exp\left(j \cdot l \cdot \frac{\pi}{2}\right)$$
$$+ \frac{A}{\sqrt{2\pi}} \cdot \exp(-j\theta) \cdot \sum_{l=-\infty}^{\infty} j^{-l} \cdot J_l(\hat{\tau}_c) \cdot \delta(\omega - (-\omega_x + l\omega_s)) \cdot \exp\left(j \cdot l \cdot \frac{\pi}{2}\right).$$

**Comment**

By going through the derivation of the analytical expression for the BPM reading of multi-bunch average, the direct influence from the number of bunches to be averaged, as shown in eq.7 is much more clear than that from the numerical calculation.